\documentclass[reqno, 12pt]{amsart}

\usepackage{_pascal/_pascal}

\newcommand{\Q}{\mathbb{Q}}
\newcommand{\R}{\mathbb{R}}
\newcommand{\Z}{\mathbb{Z}}

\newcommand{\zero}{\mathbf{0}}
\newcommand{\one}{\mathbf{1}}
\newcommand{\mim}{\operatorname{mim}}
\newcommand{\conv}{\operatorname{conv}}
\newcommand{\alignatquad}{\hspace{\minalignsep}\qquad}

\title{Combinatorial optimization with bounded subdeterminants}

\author[Fiorini]{Samuel Fiorini$^1$}
\address{$^1$Université libre de Bruxelles, Brussels, Belgium}
\email{samuel.fiorini@ulb.be}

\author[Gollin]{J.~Pascal Gollin$^2$}
\address{$^2$University of Primorska, Koper, Slovenia.}
\email{pascal.gollin@famnit.upr.si}

\author[Huynh]{Tony Huynh$^3$}
\address{$^3$Discrete Mathematics Group,  Institute for Basic Science, Daejeon, South Korea.}
\email{tony.bourbaki@gmail.com}

\author[Keil]{Simon Keil$^4$}
\address{$^4$Technical University of Munich, Munich, Germany.}
\email{simon.keil@tum.de}

\author[Kwon]{O-joung Kwon$^{5,3}$}
\address{$^5$Hanyang University, Seoul, South Korea.}
\email{ojoungkwon@hanyang.ac.kr}

\author[Oum]{Sang-il Oum$^{3,6}$}
\address{$^6$KAIST, Daejeon, South Korea.}
\email{sangil@ibs.re.kr}

\author[Sharma]{Roohani Sharma$^3$}
\email{roohani@ibs.re.kr}

\author[Weltge]{Stefan Weltge$^4$}
\email{weltge@tum.de}

\date{\today}

\subjclass[2020]{}

\begin{document}

\begin{abstract}
We take a first step toward a broader study of classical combinatorial optimization problems through the maximum subdeterminants of their standard integer programming formulations, yielding new positive results for several famous (NP-hard) problems.
An integer matrix is totally $\Delta$-modular if every square submatrix has determinant of absolute value at most $\Delta$. 
For every fixed $\Delta$, it has been known that maximum clique, maximum stable set, and minimum vertex cover can be solved in polynomial time on instances for which the coefficient matrix of the natural formulation is totally $\Delta$-modular. We extend this list by showing that the same holds for many other problems, including the weighted versions of maximum cut, the traveling salesman problem, feedback vertex set, odd cycle transversal, dominating set, max 2-sat, induced matching, and cluster vertex deletion, as well as for $k$-colorability and graph isomorphism.

Our proofs reveal several new ways in which bounded subdeterminants impose combinatorial structure: they bound the size or maximum degree of the input graph, bound packings of cycles and other obstructions, or yield decompositions into tractable graph classes such as forests, chordal graphs, graphs of bounded treewidth, and graphs of bounded linear mim-width.
\end{abstract}

\maketitle

\section{Introduction}
An integer matrix is totally $\Delta$-modular if the absolute value of the determinant of every square submatrix is at most $\Delta$.
It is a fundamental open problem whether integer programs
\[
    \max \{ c^\intercal x : Ax \le b, \, x \in \Z^n \}
\]
with a totally $\Delta$-modular coefficient matrix $A$ can be solved in polynomial time for fixed $\Delta$.
For $\Delta=1$, i.e., for totally unimodular matrices, the answer is well-known to be affirmative.
Artmann, Weismantel, and Zenklusen~\cite{artmann2017strongly} showed that polynomial-time solvability continues to hold for $\Delta=2$.
In earlier work, Shevchenko~\cite{Shevchenko1997} explicitly conjectured that the problem is polynomial-time solvable for every fixed $\Delta$.
However, the complexity status remains open for all $\Delta \ge 3$ and it is unclear which algorithmic paradigm might be suitable for solving these instances.
Instead of focusing on all instances for some fixed $\Delta$, we follow a complementary, problem-oriented viewpoint and propose a broader study of classical combinatorial optimization problems through the maximum subdeterminants of their standard integer programming formulations.

Among the list of classical combinatorial optimization problems, a first natural candidate for this study is the knapsack problem, which has an immediate integer programming formulation.
This formulation is totally $\Delta$-modular if and only if the largest size of an item is at most $\Delta$, in which case the problem is well-known to be solvable in polynomial time by dynamic programming.
While this provides a simple sanity check for the perspective we advocate, it is certainly limited in insight.

Moving beyond single constraints, graph-theoretic problems provide a significantly richer landscape for this perspective.
Natural candidates are the stable set, vertex cover, and max clique problems, which have indeed been investigated in this context.
For these problems, total $\Delta$-modularity turns out to be equivalent to bounding the odd cycle packing number of the underlying graph (or its complement) by $\log_2 \Delta$.
Initiated by Bock, Faenza, Moldenhauer, and Ruiz-Vargas~\cite{BFMR14}, a series of works~\cite{ocpgenus,ocp1,MSW21,fiorini2025integer} has shown that the weighted versions of these problems are actually polynomially solvable for any fixed $\Delta$.
Moreover, polynomial-time algorithms have been established for totally $\Delta$-modular instances of the total matching problem by Ferrarini, Fiorini, Kober, and Yuditsky~\cite{ferrarini2024total} and the unweighted dominating set problem by Malyshev and Gribanov~\cite{MG2018}.
In this work, we extend the latter result to weighted instances and give polynomial-time algorithms for several other problems.

\begin{theorem}
    \label{thmMain}
    For every constant $\Delta$, the following problems are polynomially solvable when restricted to instances for which the coefficient matrix of their natural integer programming formulation is totally $\Delta$-modular:
    \begin{itemize}
        \item $k$-colorability
        \item graph isomorphism
        \item feedback vertex set
        \item odd cycle transversal
        \item dominating set
        \item induced matching
        \item cluster vertex deletion
        \item traveling salesman problem
        \item 2-constraint satisfaction problem
    \end{itemize}
\end{theorem}

Except for $k$-colorability and graph isomorphism, we refer to the weighted optimization versions of the above problems.
By the 2-constraint satisfaction problem (max 2-CSP) we mean its Boolean version, which includes maximum cut, maximum directed cut, and max 2-SAT.
As the above result depends on the actual representations of the problems as integer programs, we explicitly state the most common formulations for each problem in the respective section of this paper.

The purpose of this work is to illustrate various new ways in which subdeterminants are linked to structural graph parameters and to highlight different algorithmic approaches needed to exploit the arising structure.
Specifically, we will see that total $\Delta$-modularity can force the input graph to be small or of bounded degree, bound the size of a packing of cycles or other obstructions, and admit decompositions into tractable graph classes such as forests, chordal graphs, graphs of bounded treewidth, and graphs of bounded linear mim-width.
We view the diversity of our results as evidence that, should a general polynomial-time algorithm for integer programs with bounded subdeterminants exist, it is unlikely to arise from a single black-box theorem in integer programming.
Instead, it may have to bring together a wide range of combinatorial arguments and algorithmic techniques.

In fact, we believe that a broader study of classical (hard) combinatorial optimization problems through the maximum subdeterminants of their integer programming formulations gives valuable insights for Shevchenko's conjecture.
Beyond the problems addressed in this work, there are many other problems whose complexity status under bounded subdeterminants remains open and which we believe deserve further investigation.
A particular example is the uncapacitated facility location problem (where the client-facility graph is not complete) as there is a direct reduction from vertex cover that preserves maximum total $\Delta$-modularity.
As the minimum cluster vertex deletion problem can be phrased as a particular covering problem whose coefficient matrix has three nonzeros per row (see \zcref{sec:clustervertexdel}), exploring other covering problems with three nonzeros per row (e.g., hitting all triangles of a graph) presents a promising direction.
Another natural question is whether our result can be extended to $k$-CSPs for $k \ge 3$, or at least 3-SAT.

It would also be interesting to explore algorithmic implications of bounded subdeterminants for problems beyond the traditional scope of combinatorial optimization.
First steps in this direction have been taken by Kuhlmann and Weismantel~\cite{kuhlmann2026threshold} who show that the $\ell_1$-shortest lattice vector problem can be solved in polynomial time for totally $\Delta$-modular instances.

The remainder of this paper is devoted to the proof of \zcref{thmMain} by treating the above problems in separate sections.

\section{$k$-colorability}\label{sec:colorability}
This problem asks whether, for a given undirected graph $G = (V, E)$ and a positive integer $k$, there exists a coloring $c: V \to [k]$ such that $c(u) \neq c(v)$ for all $\{u, v\} \in E$.
The problem is trivial for $k=1$, and for $k=2$ it is equivalent to checking whether $G$ is bipartite.
From now on, assume $k \ge 3$, for which the problem is NP-complete.
A typical formulation of the problem as a (feasibility) integer program is
\begin{equation}
    \label{eqipcolor}
    \begin{aligned}
        & x_{v, i} + x_{u, i} \leq 1 && \qquad\forall \{u, v\} \in E, i \in [k] \\
        & \sum_{i=1}^k x_{v, i} = 1 &&  \qquad\forall v \in V \\
        & x_{v, i} \in \{0, 1\} &&  \qquad\forall v \in V, i \in [k]
    \end{aligned}
\end{equation}
where the variable $x_{v, i}$ indicates whether vertex $v$ is colored with color $i$ for each $v \in V$ and $i \in [k]$.
Under the assumption of bounded subdeterminants, we only have to consider input graphs with the following property.
\begin{lemma}
\label{lemcolor}
If the coefficient matrix of~\eqref{eqipcolor} is totally $\Delta$-modular, then $G$ contains at most $\lfloor \log_2\Delta \rfloor$ vertex disjoint cycles.
\end{lemma}
\begin{proof}
Let $B \in \{0, 1\}^{E \times V}$ be the edge-vertex-incidence matrix of $G$. That is, $B_{\{u, v\}, w} = 1$ if $w \in \{u, v\}$ and $0$ otherwise.  Collect the variables into one vector $x = (x^{(1)}, \dots, x^{(k)})$ where $x^{(i)} = (x_{v, i})_{v \in V}$ for $i \in [k]$.
Then we can rewrite~\eqref{eqipcolor} as
\begin{align*}
    \begin{pmatrix}
        B & \zero & \dots & \zero \\
        \zero & B & \dots & \zero \\
        \vdots & \vdots & \ddots & \vdots \\
        \zero & \zero & \dots & B \\
    \end{pmatrix} x &\leq \one \\
    \begin{pmatrix}
        I & I & \dots & I \\
    \end{pmatrix}x &= \one \\
    x & \in \{0, 1\}^{V \times [k]}.
\end{align*}
Let $C$ be an arbitrary cycle of $G$ with vertices $v_1, \dots, v_\ell$.
If $C$ has odd length, then the corresponding submatrix of $B$ (say in the first color) has determinant $\pm 2$.
If $C$ has even length, we find the incidence matrix of a cylce of length $\ell + 3$ as a submatrix of the coefficient matrix of~\eqref{eqipcolor}. As $\ell + 3$ is odd, the resulting submatrix will have determinant of absolute value $2$. To this end, select the columns corresponding to $x_{v_1, 1}, x_{v_2, 1}, x_{v_2, 2}, x_{v_3, 2}, x_{v_3, 3}, x_{v_4, 3}, x_{v_4, 1}, \dots, x_{v_\ell, 1}$ and the rows corresponding to $\{v_1, v_2\}$ in the first color, $\{v_2, v_3\}$ in the second color, $\{v_3, v_4\}$ in the third color, $\{v_i, v_{i+1}\}$ in the first color for $i = 4, \dots, \ell - 1$, as well as the rows corresponding to $v_2, v_3, v_4$ in the matrix $\begin{pmatrix} I & I & \dots & I \end{pmatrix}$. The construction is illustrated in \zcref{fig:coloring-even-cycle}.
\begin{figure}[ht]
    \centering
    \includegraphics{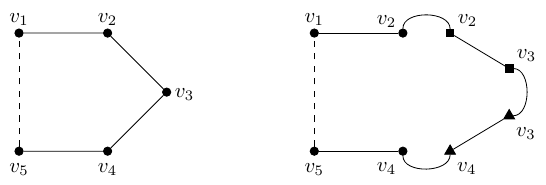}
    \caption{Illustration of the proof of \zcref{lemcolor}: An even cycle in the original graph (left) and a corresponding odd cycle whose incidence matrix we find as a submatrix of the coefficient matrix (right). The dashed edge represents a path of even length (possibly zero, in which case $v_1 = v_5$). The curved edges represent the edges corresponding to the rows of the matrix $\begin{pmatrix} I & I & \dots & I \end{pmatrix}$. The shape of the vertices on the right together with the labels indicate the respective chosen columns.}
    \label{fig:coloring-even-cycle}
\end{figure}
If $C_1,\dots,C_t$ are vertex-disjoint cycles, we can apply the above construction to each of these cycles.
Since the rows selected for any of the cycles have 0 entries in the columns selected for the other cycles, the determinant of the resulting submatrix is the product of the determinants of the submatrices corresponding to the individual cycles and hence $\pm 2^t$.
\end{proof}

In view of \zcref{thmMain}, we can exploit the above statement as follows.
By the Erdős-Pósa theorem~\cite{ErdosP1965}, there is a set $U$ of $O(\log\Delta\log\log\Delta)$ vertices that hits all cycles.
For constant $\Delta$, we can efficiently enumerate all possible choices and potential colorings of $U$.
For each of these choices, the remaining problem is a list coloring problem (a coloring problem in which each vertex has a list of allowed colors) in a forest. This can be solved in linear time by an algorithm due to Jansen and Scheffler~\cite[\S 3]{jansen1997generalized}.

\section{Graph isomorphism}
The graph isomorphism problem asks whether two given graphs are equal up to a relabeling of the vertices. In other words, is there a bijection between the vertex sets of the two graphs that preserves adjacency and nonadjacency?
Graph isomorphism is not known to be solvable in polynomial time nor known to be NP-complete.
Babai~\cite{babai2016graph} showed that it can be solved in quasipolynomial time.

If $G$ and $G'$ are undirected graphs with adjacency matrices $A$ and $A'$, respectively, note that $G$ and $G'$ are isomorphic if and only if there exists a permutation matrix $P$ with $A = P A' P^\intercal$, or, equivalently, $AP = PA'$.
This gives rise to the following common (feasibility) integer programming formulation of the problem.
\begin{alignat}{2}
    &A X = X A' \label{eq:iso-cons}\\
    &\sum_{u \in V} x_{uv} = 1 && \alignatquad\forall v \in V \nonumber \\
    &\sum_{v \in V} x_{uv} = 1 && \alignatquad\forall u \in V \nonumber \\
    &x_{uv} \in \{0, 1\} && \alignatquad\forall u, v \in V \nonumber 
\end{alignat}
\begin{lemma}
\label{lem:iso-subdet}
If the coefficient matrix of~\eqref{eq:iso-cons} is totally $\Delta$-modular and $G'$ has at least one edge, then the maximum degree of $G$ is at most $\Delta + 1$.
\end{lemma}
\begin{proof}
We may assume that $G$ and $G'$ have the same vertex set $V$.
For each $(u,v) \in V^2$, comparing the respective entries of $AX$ and $XA'$, the equation in~\eqref{eq:iso-cons} reads
\[
    \sum_{w \in N(u)} x_{wv} - \sum_{w \in N'(v)} x_{uw} = 0,
\]
where $N(w)$ and $N'(w)$ denote the sets of neighbors of $w$ in $G$ and $G'$, respectively.
Treating $x$ as a vector in $\{0,1\}^{V^2}$ we can rewrite the above constraints as $Bx = \zero$, where $B \in \{-1, 0, 1\}^{V^2 \times V^2}$ is given by
\[
    B_{(u, v), (w, z)} = \begin{cases}
        +1 & \text{if $z = v$ and $w \in N(u)$}, \\
        -1 & \text{if  $w = u$ and $z \in N'(v)$}, \\
        \phantom{+}0 & \text{otherwise}.
    \end{cases}
\]
By the assumption, $B$ is totally $\Delta$-modular.
Let $u$ be a vertex in $G$ with neighbors $v_1,\dots,v_d$ in $G$, and let $w$ denote a vertex in $G'$ that has at least one neighbor $z$ in $G'$.
Consider the submatrix of $B$ that consists of the rows corresponding to $(u, w)$ and $(v_i, z)$ and the columns corresponding to $(u, z)$ and $(v_i, w)$ for $i = 1,\dots,d$.
The form of this submatrix is
\[
    D = \begin{pmatrix}
        -1 & 1 & \cdots & 1 \\
        1 & -1 & & \\
        \vdots & & \ddots & \\
        1 &  & & -1
    \end{pmatrix}
\]
where all unspecified entries are zero.
By adding all rows to the first row, we obtain $d - 1 = |\det D| \le \Delta$, which yields the claim.
\end{proof}
\zcref{lem:iso-subdet} allows us to reduce the problem to the isomorphism problem for graphs of bounded degree, which was shown to be polynomially solvable by Luks~\cite{luks1982isomorphism}.

\section{Feedback vertex set and odd cycle transversal}
The feedback vertex set problem asks to find a minimum-weight set of vertices that intersects every cycle in a given graph, or equivalently, such that the graph obtained by deleting these vertices is a forest. Similarly, the odd cycle transversal problem asks to find a minimum-weight set of vertices that intersects every odd cycle in a given graph, or equivalently, such that the graph obtained by deleting these vertices is bipartite. 

Given a graph $G = (V, E)$ and a weight function $w: V \to \Q$, a straightforward integer programming formulation of the problems is
\begin{equation}
    \label{eq:hitting-set}
    \begin{aligned}
        \min\ & \sum_{v \in V} w(v) x_v \\
        \text{s.\,t.}\ & \sum_{v \in V(C)} x_v \geq 1 && \qquad\forall C \in \mathcal{C} \\
        & x_v \in \{0, 1\} && \qquad\forall v \in V
    \end{aligned}
\end{equation}
where $\mathcal{C}$ is the set of all cycles for the feedback vertex set problem, respectively, the set of all odd cycles for the odd cycle transversal problem, and $x_v$ indicates whether the vertex $v$ is included in the solution.

We remark that, for fixed $\Delta$, it is known that a totally $\Delta$-modular matrix with $n$ columns has only polynomially many (in $n$) distinct rows (see~\cite{paat2024column}). Hence, in any totally $\Delta$-modular instance, \eqref{eq:hitting-set} has only polynomially many distinct constraints. However, we do not see how this alone implies that the problem can be solved in polynomial time. Instead, we will use that restricting feedback vertex set instances to those with bounded subdeterminants excludes large grid minors, which allows us to bound the treewidth of the input graph. We then extend this approach to the odd cycle transversal problem.

\subsection{Feedback vertex set}
To formalize the grid minor argument, let us first fix the minor terminology we use. Given graphs $G$ and $H$, an \emph{$H$-model} $M$ in $G$ consists of a non-empty tree $M(v)$ that is a subgraph of $G$ for each vertex $v \in V(H)$, called a \emph{branch set}, and an edge $M(\{u,v\}) \in E(G)$ with one endpoint in $M(u)$ and the other in $M(v)$ for each edge $\{u,v\} \in E(H)$, such that all branch sets are pairwise vertex-disjoint. $G$ has an \emph{$H$-minor} if and only if there is an $H$-model in $G$.

\begin{lemma}
    \label{lem:fvs-grid-minor}
    For each $k \ge 2$, if a graph $G$ contains a $(2k \times 2k)$-grid as a minor, then the coefficient matrix of \eqref{eq:hitting-set} with $\mathcal{C}$ being the set of all cycles in $G$ contains every non-singular $0/1$-matrix of size $k \times k$ as a submatrix.
\end{lemma}
\begin{proof}
    Let $A$ be an arbitrary non-singular $0/1$-matrix of size $k \times k$. In particular, all rows of $A$ are distinct. We will first describe the case where $G$ contains a $(2k \times 2k)$-grid as a subgraph, not only as a minor. Label the vertices of the grid as $v_{i,j}$ for $i,j \in [2k]$ where $i$ is the index of the grid row and $j$ is the index of the grid column. We build our submatrix by selecting the columns corresponding to vertices in the first grid row with odd grid column index, that is, $v_{1,2\ell-1}$ for $\ell \in [k]$.
    Then, for each row $a \in \{0,1\}^k$ of the matrix $A$, we construct a cycle that yields precisely this row in the submatrix. To this end, choose any path from $v_{2,1}$ to $v_{2,2k}$ that intersects the first two grid rows only in its endpoints. Then, for each entry $a_i$ of $a$ for $i \in [k-1]$, construct a path from $v_{2,2i-1}$ to $v_{2,2i+1}$ as follows. If $a_i = 0$, use the path joining $v_{2,2i-1}$ and $v_{2,2i+1}$ in the second grid row, thus avoiding $v_{1,2i-1}$. If $a_i = 1$, use the path $v_{2,2i-1}, v_{1,2i-1}, v_{1,2i}, v_{2,2i}, v_{2,2i+1}$, thus using $v_{1,2i-1}$. For $a_k$, construct a path from $v_{2,2k-1}$ to $v_{2,2k}$ by using an analogous construction. By joining all these paths, we obtain a cycle that intersects the selected vertices in the first grid row precisely in the vertices corresponding to the 1-entries of $a$.

    Now, assume that $G$ has a $(2k \times 2k)$-grid $H$ as a minor. Let $M$ be an $H$-model in $G$. Consider the selection of vertices and cycles in $H$ described above. For each selected cycle in $H$, we obtain a corresponding cycle in $G$ by taking $M(e)$ for each edge $e$ of the cycle and, for each vertex $v$ of the cycle, connecting the relevant endpoints by a path in $M(v)$. Concerning the selected vertices, some care is needed, since the branch set $M(v)$ of a vertex $v \in V(H)$ may contain multiple vertices of $G$, some of which might not be part of all lifted cycles through $M(v)$. However, note that the selected vertices in $H$ all have degree at most 3. A helpful fact is that in any tree and for any three (not necessarily distinct) vertices of that tree, there is a unique vertex that is contained in all three paths between any two of the three vertices. Using this fact, we can select a vertex $v' \in M(v)$ for each selected vertex $v$ that is contained in the selected cycles in $G$ if and only if $v$ is contained in the corresponding cycles in $H$.
\end{proof}

Note that there are $0/1$-matrices of size $k \times k$ with determinant exponential in $k$. Hence, by the above lemma, any graph for which the formulation \eqref{eq:hitting-set} is totally $\Delta$-modular cannot contain a large grid as a minor. The grid minor theorem of Robertson and Seymour~\cite{robertson1986graph} states that such graphs have bounded treewidth (see~\cite[\S 12]{diestel2025graph} for a definition of treewidth).

More concretely, we conclude that there exists a function $f$ such that the treewidth of every graph $G$, for which the formulation \eqref{eq:hitting-set} is totally $\Delta$-modular, is at most $f(\Delta)$. Moreover, the feedback vertex set problem can be solved in polynomial time on graphs of bounded treewidth (see~\cite{fomin2017representative}). This shows that, for each fixed $\Delta$, totally $\Delta$-modular instances of the feedback vertex set problem can be solved in polynomial time.

\subsection{Odd cycle transversal}
In order to show the corresponding result for the odd cycle transversal problem, we adapt the approach for the feedback vertex set problem using the techniques of~\cite{gollin2023structure}. Namely, we replace minors and grids by suitable variants to control the parity of cycles.

Let $M$ be an $H$-model in a graph $G$. We call $M$ an \emph{odd $H$-model} if the vertices of its branch sets admit a $2$-coloring that is proper on each branch set $M(v)$ and under which each edge $M(e)$ is monochromatic. Accordingly, we say that $H$ is an \emph{odd minor} of $G$ if $G$ contains an odd $H$-model. It is easy to see that odd minors preserve the parity of cycles in the sense that every cycle in $H$ gives rise to a cycle of the same parity in $G$ with the natural lifting described in the proof of \zcref{lem:fvs-grid-minor}.

\begin{figure}[htbp]
    \centering
    \includegraphics{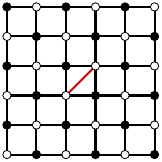}
    \caption{The singly parity-breaking grid of order~$3$.}
    \label{fig:paritybreakinggrid}
\end{figure}

The role of grids in the case of the feedback vertex set problem is taken over by \emph{singly parity-breaking grids}. We define such a grid of \emph{order~$k$} as the graph obtained from the ${(2k\times 2k)}$-grid by adding a single parity-breaking edge in the center, see \zcref{fig:paritybreakinggrid} for an illustration of the case $k=3$.

We obtain the following lemma which corresponds to \zcref{lem:fvs-grid-minor}.
\begin{lemma}
    \label{lem:oct-odd-minor}
    For each $k \ge 3$, if a graph $G$ contains a singly parity-breaking grid of order $k$ as an odd minor, then the coefficient matrix of \eqref{eq:hitting-set} with $\mathcal{C}$ being the set of all odd cycles in $G$ contains every non-singular $0/1$-matrix of size $k \times k$ as a submatrix.
\end{lemma}
\begin{proof}
    The proof is completely analogous to the proof of \zcref{lem:fvs-grid-minor}. The only difference is that we choose the path from $v_{2,1}$ to $v_{2,2k}$ such that it uses the parity-breaking edge exactly once, making the cycle odd. This is possible since the parity-breaking edge lies outside the first two rows for $k \ge 3$.
\end{proof}

Since there are $0/1$-matrices with determinant exponential in their size, \zcref{lem:oct-odd-minor} implies that if the coefficient matrix of \eqref{eq:hitting-set} for odd cycle transversal is totally $\Delta$-modular, then the underlying graph does not contain a singly parity-breaking grid of order $k(\Delta)$ for a function $k(\Delta) = O(\log \Delta)$ as an odd minor.

Graphs without such an odd minor are well-structured in terms of their blocks. To make this precise, define a \emph{cutvertex} of a graph to be a vertex whose removal increases the number of connected components and a \emph{block} of a graph as a maximal connected subgraph without a cutvertex. Note that a block is either a maximal $2$-connected subgraph, a single edge, or an isolated vertex. By a classical result of Tutte~\cite{tutte1966} the graph $T_{bc}(G)$ that has one vertex for each block and each cutvertex of a connected graph~$G$, where a block $X$ is adjacent to a cutvertex $v$ if $v \in V(X)$, is a tree, called the \emph{block-cut tree} of~$G$. $T_{bc}(G)$ can be computed in linear time (see~\cite{Schmidt2013}). Note that every cycle is contained in a single block, and hence a vertex set is an odd cycle transversal of $G$ if and only if its intersection with each block is an odd cycle transversal of that block.

\begin{lemma}
    \label{lem:oct-decomposition}
    There is a function $g$ such that, for every positive integer $k$, every block of a graph that does not contain the singly parity-breaking grid of order~$k$ as an odd minor is bipartite or has treewidth at most $g(k)$.
\end{lemma}
\begin{proof}
    Blocks that are not $2$-connected are isomorphic to $K_1$ or $K_2$ and hence bipartite. Since an odd model in a subgraph is also an odd model in the whole graph, the claim follows from~\cite[Lemma~5.6]{gollin2023structure}, which implies that every non-bipartite $2$-connected graph without the singly parity-breaking grid of order~$k$ as an odd minor has treewidth at most $g(k)$.
\end{proof}

We can now give a dynamic programming algorithm similar to~\cite[\S 7]{gollin2023structure}.

\begin{lemma}
    \label{lem:oct-dp}
    Let $\ell$ be a fixed integer. For a graph $G$ each of whose blocks is bipartite or has treewidth at most~$\ell$, a minimum-weight odd cycle transversal can be computed in polynomial time.
\end{lemma}

\begin{proof}
    On a single block, the problem can be solved in polynomial time. For bipartite graphs, the problem is trivial and for graphs of bounded treewidth the polynomial-time algorithm developed in \cite[\S 3]{fiorini2008planar} can directly be extended to the weighted case.

    We may assume that $G$ is connected and has a cutvertex, as otherwise we can treat the components separately or $G$ is a single block. We compute the block-cut tree $T_{bc}(G)$ and root it at a cutvertex~$r$. Then each block $X$ has a parent $p_X \in V(X)$ in $T_{bc}(G)$, and its children $C_X$ are the cutvertices in $V(X) \setminus \{p_X\}$. The children $C_v$ of a cutvertex $v$ are the blocks $X$ with $p_X = v$. For a cutvertex $v$, let $G_v$ be the union of all blocks in the subtree of $T_{bc}(G)$ rooted at $v$. In particular, we have $G_r = G$. Let $\mathrm{opt}_v(0)$ be the minimum weight of an odd cycle transversal of $G_v$ that does not contain $v$, and let $\mathrm{opt}_v(1)$ be the minimum weight of an odd cycle transversal of $G_v$ that contains $v$. We compute these values bottom-up in $T_{bc}(G)$ and return $\min\{\mathrm{opt}_r(0), \mathrm{opt}_r(1)\}$.
 
    Let $v$ be a cutvertex. For each child $X$ of $v$, define weights on $V(X) \setminus \{v\}$ by $w_X(x) = \mathrm{opt}_x(1) - \mathrm{opt}_x(0)$ for $x \in C_X$ and $w_X(x) = w(x)$ otherwise. We claim that for $b \in \{0, 1\}$, we have
    \[
        \mathrm{opt}_v(b) = b \cdot w(v) + \sum_{X \in C_v} \Bigl(\min \, \bigl\{w_X(S \setminus \{v\}): S \in \mathcal{T}_X, \chi_S(v) = b\bigr\} + \sum_{x \in C_X} \mathrm{opt}_x(0)\Bigr),
    \]
    where $\mathcal{T}_X$ denotes the set of all odd cycle transversals of $X$ and $\chi_S$ denotes the characteristic vector of $S$, i.e., $\chi_S(x) = 1$ if $x \in S$ and $\chi_S(x) = 0$ otherwise. To see the correctness, note that the blocks of $G_v$ are the children of $v$ and the blocks of the graphs $G_x$. Hence, a set $S \subseteq V(G_v)$ is an odd cycle transversal of $G_v$ if and only if $S \cap V(X)$ is one of $X$ and $S \cap V(G_x)$ is one of $G_x$ for all $X \in C_v$ and $x \in C_X$. Further, $V(G_v)$ is the disjoint union of $\{v\}$ and, for each $X \in C_v$, the set $V(X) \setminus (C_X \cup \{v\})$ and the sets $V(G_x)$ with $x \in C_X$, and $V(G_x) \cap V(X) = \{x\}$. Hence, given $\chi_S(v) = b$, the sets $S \cap V(X)$ can be chosen independently for the children $X$ of $v$, and once they are fixed, so can the sets $S \cap V(G_x)$, each optimally at cost $\mathrm{opt}_x(\chi_S(x)) = \mathrm{opt}_x(0) + \chi_S(x) \cdot w_X(x)$. This proves the claim.
    Each of these minima can be computed by a weighted odd cycle transversal problem on $X$, as the constraint $\chi_S(v) = b$ can be dropped by setting $w_X(v) = -1$ if $b = 1$, so that adding $v$ to an odd cycle transversal not containing it yields one of smaller weight, and $w_X(v) = 1 + \sum_{x \in N_X(v)} \max\{0, w_X(x)\}$ if $b = 0$, since then replacing $v$ by its neighbors in $X$ turns any odd cycle transversal containing $v$ into one of smaller weight. Thus, we obtain all values $\mathrm{opt}_v(b)$ by solving two odd cycle transversal problems per block. An optimal solution can be recovered by standard backtracking.
\end{proof}

By \zcref{lem:oct-odd-minor} and the remark following it, and \zcref{lem:oct-decomposition,lem:oct-dp}, we conclude that, for each fixed $\Delta$, instances of the odd cycle transversal problem for which the coefficient matrix of~\eqref{eq:hitting-set} is totally $\Delta$-modular can be solved in polynomial time.

An alternative common integer programming formulation of the odd cycle transversal problem is given by
\begin{equation}
    \label{eq:oct-bipartite}
    \begin{aligned}
        \min\ & \sum_{v \in V} w(v) x_{vd} \\
        \text{s.\,t.}\ & x_{vl} + x_{vr} + x_{vd} = 1 && \qquad\forall v \in V \\
        & x_{vl} + x_{ul} \leq 1 && \qquad\forall \{u, v\} \in E\\
        & x_{vr} + x_{ur} \leq 1 && \qquad\forall \{u, v\} \in E\\
        & x_{vl}, x_{vr}, x_{vd} \in \{0, 1\} && \qquad\forall v \in V
    \end{aligned}
\end{equation}
where $x_{vl}$, $x_{vr}$ and $x_{vd}$ indicate whether $v$ is assigned to the left side, assigned to the right side, or deleted, respectively.

Let $\Delta$ be fixed.
We show that every instance of the odd cycle transversal problem for which the coefficient matrix of \eqref{eq:oct-bipartite} is totally $\Delta$-modular can be also solved in polynomial time.
To this end, we make use of a result by Fiorini, Joret, Weltge, and Yuditsky~\cite{fiorini2025integer} stating that totally $\Delta$-modular integer programs with at most two variables per inequality can be solved in polynomial time.
Note that the equality constraints in the above formulation imply $x_{vd} = 1 - x_{vl} - x_{vr}$ for each $v \in V$.
Substituting this in the objective and all constraints (including the bounds) yields an equivalent integer program with at most two variables per inequality.
Its coefficient matrix is a submatrix of the coefficient matrix of \eqref{eq:oct-bipartite}, so it is totally $\Delta$-modular if \eqref{eq:oct-bipartite} is.
Hence, we can apply~\cite{fiorini2025integer} to obtain the claimed polynomial-time algorithm.

Interestingly, the two different formulations for the odd cycle transversal problem yield two incomparable graph classes when restricted to bounded subdeterminants. On the one hand, the disjoint union of~$k$ odd cycles has a totally unimodular coefficient matrix with respect to \eqref{eq:hitting-set}, but has a subdeterminant of absolute value~$4^k$ with respect to \eqref{eq:oct-bipartite}. On the other hand, the singly parity-breaking grid of order~$k$ has a totally 4-modular coefficient matrix with respect to \eqref{eq:oct-bipartite}, but has subdeterminants of absolute value exponential in~$k$ with respect to \eqref{eq:hitting-set}. This can be seen by \zcref{lem:oct-odd-minor} and the fact that the largest absolute value of a subdeterminant of the edge-vertex incidence matrix of a graph is equal to~$2^k$, where~$k$ is the maximum number of vertex-disjoint odd cycles in the graph \cite{grossman1995minors}.

\section{Dominating set}
Given a graph $G=(V,E)$ and a weight function $w: V \to \Q$, the dominating set problem asks to find a subset $D \subseteq V$ of minimum weight such that every vertex in $V \setminus D$ has a neighbor in $D$.
A natural integer programming formulation of the dominating set problem is given by
\begin{equation}
    \label{eq:dominating-set-ip}
    \begin{aligned}
        \min\ & \sum_{v \in V} w(v) x_v \\
        \text{s.\,t.}\ & x_v + \sum_{u \in N(v)} x_u \geq 1 && \qquad\forall v \in V\\
        & x_v \in \{0, 1\} && \qquad\forall v \in V
    \end{aligned}
\end{equation}
where $N(v)$ denotes the set of neighbors of $v$ in $G$.

\begin{figure}[hbt]
    \centering
    \begin{subfigure}[b]{0.25\textwidth}
        \centering
        \includegraphics{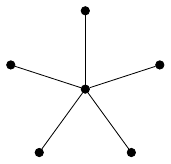}
        \caption{$5$-star}
    \end{subfigure}
    \hspace{1cm}
    \begin{subfigure}[b]{0.25\textwidth}
        \centering
        \includegraphics{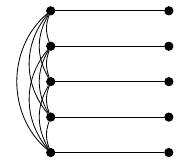}
        \caption{$5$-net}
    \end{subfigure}
    \hspace{1cm}
    \begin{subfigure}[b]{0.25\textwidth}
        \centering
        \includegraphics{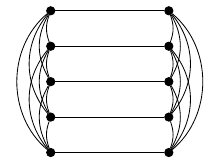}
        \caption{$5$-prism}
    \end{subfigure}
    \caption{The $k$-star, $k$-net, and $k$-prism for $k=5$.}
    \label{fig:starnetprism}
\end{figure}

We start by fixing some notation needed for our argument. The \emph{$k$-star} is the complete bipartite graph with one part of size one and the other of size $k$, the \emph{$k$-net} is the graph consisting of a clique of size $k$ and a disjoint independent set of size $k$ joined to the clique by a perfect matching, and the \emph{$k$-prism} is a graph consisting of two disjoint cliques of size $k$ joined by a perfect matching. See \zcref{fig:starnetprism} for an illustration. The $3$-star is also known as the \emph{claw}, and the $3$-net as the \emph{net}.

For a collection of graphs $\mathcal{H}$, we say that a graph $G$ is \emph{$\mathcal{H}$-free} if it does not contain any graph from $\mathcal{H}$ as an induced subgraph. Let $\mathcal{H}_\Delta$ be the collection consisting of the $(\Delta + 2)$-star, $(\Delta + 2)$-net, $(\Delta + 1)$-prism, the disjoint union of $t_\Delta$ claws, and the disjoint union of $t_\Delta$ nets, where $t_\Delta = \lceil \log_2 \Delta \rceil + 1$.

In the unweighted setting, Malyshev and Gribanov \cite{MG2018} proved polynomial-time solvability by showing that, after guessing a number of vertices depending only on $\Delta$, the problem reduces to an instance on \{claw, net\}-free graphs. The latter problem was shown to be solvable in polynomial time by Brandstädt and Dragan \cite{BD2003}. However, as the following proposition shows, this approach does not extend to the weighted case.

\begin{proposition}
    The weighted version of the dominating set problem on \{claw, net\}-free graphs is NP-hard.    
\end{proposition}
\begin{proof}
    Recall the NP-hard set cover problem: Given a set $U$ and a family $\mathcal{F}$ of subsets of $U$, the set cover problem asks to find a minimum-size subfamily $\mathcal{F}' \subseteq \mathcal{F}$ such that every element in $U$ is contained in at least one set in $\mathcal{F}'$.

    To reduce an instance of set cover to a \{claw, net\}-free instance of weighted dominating set, create a graph $G$ with vertex set $U \cup \mathcal{F}$, such that $G[U]$ and $G[\mathcal{F}]$ are cliques and there is an edge between $u \in U$ and $F \in \mathcal{F}$ if and only if $u \in F$. Assign each vertex in $U$ a weight of $|U|+1$ and each vertex in $\mathcal{F}$ a weight of $1$. Then a minimum weight dominating set of $G$ corresponds to a minimum size set cover of the original instance and the resulting graph is \{claw, net\}-free, since it does not contain an independent set of size more than $2$.
\end{proof}

While this shows restricting to \{claw, net\}-free graphs is not sufficient, they still play an important role in our proof. For a vertex $v$, let $N_k(v)$ denote the set of vertices at distance exactly $k$ from $v$. We use the following property of \{claw, net\}-free graphs.

\begin{lemma}
    \label{lem:dominating-claw-net-free}
    Let $H$ be a \{claw, net\}-free graph, $k\ge 0$, and $r\in V(H)$. Then $H[N_k(r)]$ contains no independent set of size at least 3.
\end{lemma}
\begin{proof}
    We prove the claim by induction on $k$. For $k = 0$, the claim is trivial. For $k = 1$, the claim follows from the fact that $H$ is claw-free.
    Let $k \ge 2$. First, consider any vertex $u \in N_{k-1}(r)$, assume that there are two non-adjacent vertices $x, y \in N(u) \cap N_k(r)$, and let $z$ be a neighbor of $u$ in $N_{k-2}(r)$. Then $\{u, x, y, z\}$ induces a claw in $H$. Therefore the set $N(u) \cap N_k(r)$ is a clique.
    Now, assume for contradiction that $H[N_k(r)]$ contains an independent set $\{v_1, v_2, v_3\}$ of size 3. Let $u_i$ be a neighbor of $v_i$ in $N_{k-1}(r)$ for $i \in \{1, 2, 3\}$. Since $N(u_i) \cap N_k(r)$ is a clique, the $u_i$ must be pairwise distinct and $u_i$ is adjacent to $v_j$ if and only if $i = j$. By the induction hypothesis, there is at least one edge among the $u_i$. Moreover, if there are exactly two edges, say $\{u_1, u_2\}$ and $\{u_2, u_3\}$, then $\{u_1, u_2, u_3, v_2\}$ induces a claw in $H$. If there are three edges, then $\{u_1, u_2, u_3, v_1, v_2, v_3\}$ induces a net in $H$. It remains to consider the case that there is exactly one edge among the $u_i$, say $\{u_1, u_2\}$. Let $w$ be a neighbor of $u_1$ in $N_{k-2}(r)$. If $w$ is not adjacent to $u_2$, then $\{w, u_1, u_2, v_1\}$ induces a claw in $H$. If $w$ is adjacent to $u_2$ and $k \geq 3$, pick a neighbor $s$ of $w$ in $N_{k-3}(r)$. Then $\{s, w, u_1, u_2, v_1, v_2\}$ induces a net in $H$. If $w$ is adjacent to $u_2$ and $k = 2$, then $u_3 \in N_1(r) = N(r)$ and $w = r$. In this case, $\{r, u_1, u_2, u_3, v_1, v_2\}$ induces a net in $H$.
\end{proof}

Moreover, Malyshev and Gribanov~\cite{MG2018} observed the following.
\begin{lemma}[{\cite[Lemma~1]{MG2018}}]
    \label{lem:dominating-forbidden-subgraphs}
    If the coefficient matrix of \eqref{eq:dominating-set-ip} is totally $\Delta$-modular, then~$G$ is $\mathcal{H}_\Delta$-free.
\end{lemma}

We will use this in combination with the notion of \emph{linear mim-width}. Let $G = (V, E)$ be a graph. For disjoint sets $A,B \subseteq V$ of vertices, let $G[A,B]$ denote the bipartite graph with vertex set $A\cup B$ and edge set $\{\{a,b\}\in E: a\in A \text{ and } b\in B\}$. For $A\subseteq V$, $\mim_G(A)$ denotes the maximum size of an induced matching in $G[A,V\setminus A]$.
A \emph{linear ordering} of $G$ is a permutation of $V$. The \emph{mim-width} of a linear ordering $(v_1,\ldots,v_n)$ of $G$ is defined as $\max_{1\leq i\leq n} \left\{ \mim_G(\{v_1,\ldots,v_i\})\right\}$.
The \emph{linear mim-width} of $G$ is the minimum mim-width over all linear orderings of $G$.

Further, we write $R(s,t)$ for the minimum integer $N$ such that every graph on $N$ vertices contains either an independent set of size $s$ or a clique of size $t$. The classical theorem of Ramsey~\cite{Ramsey1930} implies that $R(s,t)$ exists.

\begin{lemma}
    \label{lem:dominating-lmimw}
    The linear mim-width of an $\mathcal{H}_\Delta$-free graph is at most
    \[
        f(\Delta) = R(10 \lceil \log_2 \Delta \rceil (\Delta+1) + 5, R(\Delta+2, \Delta+1)) - 1.
    \]
    Moreover, a linear ordering with mim-width at most $f(\Delta)$ can be found in polynomial time.
\end{lemma}
\begin{proof}
    Let $\mathcal{P}$ be a maximal packing of induced subgraphs in $G$, each isomorphic to the claw or the net, such that there is no edge between distinct members of $\mathcal{P}$. Note that such a packing $\mathcal{P}$ can be found greedily in polynomial time. Set $P = \bigcup_{Q \in \mathcal{P}} V(Q)$. Since $G$ is $\mathcal{H}_\Delta$-free, $\mathcal{P}$ contains at most $(t_\Delta - 1)$ claws (with 4 vertices each) and at most $(t_\Delta - 1)$ nets (with 6 vertices each) and hence $|P| \leq 10 (t_\Delta - 1)$.

    Let $K_1, \dots, K_t$ be the connected components of $G[V \setminus N[P]]$. For each $i \in [t]$, let $r_i$ be an arbitrary vertex in $K_i$. Consider the linear ordering of $G$ obtained by first taking an arbitrary ordering of $N[P]$ and then appending, for each connected component $K_i$ in order of increasing index, all vertices of $K_i$ in order of non-decreasing distance from $r_i$ in $K_i$. The indexing of the connected components $K_i$ and the order of the vertices within $K_i$ with the same distance from $r_i$ can be chosen arbitrarily. Note that such an ordering can be computed in polynomial time. We claim that this ordering witnesses that $G$ has linear mim-width at most $R(|P|(\Delta + 1) + 5, R(\Delta+2, \Delta+1)) - 1 \leq R(10 \lceil \log_2 \Delta \rceil (\Delta+1) + 5, R(\Delta+2, \Delta+1)) - 1$.

    To this end, let $S$ be a prefix of the ordering and let $\{\{a_1, b_1\}, \dots, \{a_\ell, b_\ell\}\}$ denote a maximum induced matching in $G[S, V \setminus S]$ with $A = \{a_1, \dots, a_\ell\} \subseteq S$ and $B = \{b_1, \dots, b_\ell\} \subseteq V \setminus S$. Let $\alpha(Q)$ and $\omega(Q)$ denote the size of a maximum independent set and a maximum clique in a graph $Q$, respectively. To prove the claim, it suffices to show that $\alpha(G[A]) \leq |P|(\Delta + 1) + 4$ and $\omega(G[A]) \leq R(\Delta+2, \Delta+1) - 1$.

    Note that
    \[
        \alpha(G[A]) \leq \alpha(G[A \cap N[P]]) + \alpha(G[A \setminus N[P]]) \leq \sum_{v \in P} \alpha(G[N[v]]) + \alpha(G[A \setminus N[P]]).
    \]
    Since $G$ does not contain a $(\Delta + 2)$-star as an induced subgraph, $\alpha(G[N[v]]) \leq \Delta + 1$ for each $v \in P$. By the construction of the ordering, either $S \setminus N[P]$ is empty, in which case $\alpha(G[A \setminus N[P]]) = 0$, or $N[P] \subseteq S$ which means $B \subseteq V \setminus N[P]$. In the latter case, let $j^*$ be the maximal index of a connected component $K_{j^*}$ such that $A \cap V(K_{j^*}) \neq \emptyset$ and let $k^*$ be the maximal distance in $K_{j^*}$ from $r_{j^*}$ to any vertex in $A  \cap V(K_{j^*})$. Then, for any $a \in A \setminus N[P]$ and its matched vertex $b \in B$, both $a$ and $b$ must be in $K_{j^*}$ and the distance from $r_{j^*}$ to $a$ must be $k^*$ or $k^* - 1$. By the definition of $\mathcal{P}$, the graph $G[V \setminus N[P]]$ is \{claw, net\}-free. Hence, by \zcref{lem:dominating-claw-net-free}, we have that $\alpha(G[A \setminus N[P]]) \leq 2 \cdot 2$ since $A \setminus N[P]$ intersects at most two layers. Altogether, we have $\alpha(G[A]) \leq |P|(\Delta + 1) + 4$.

    To show that $\omega(G[A]) \leq R(\Delta+2, \Delta+1) - 1$, assume for contradiction that $G[A]$ contains a clique of size $R(\Delta+2, \Delta+1)$ with vertex set $A'$. Let $B'$ be the set of vertices matched to $A'$ in the induced matching. By the definition of $R(\Delta+2, \Delta+1)$, $G[B']$ contains either an independent set of size $\Delta + 2$ or a clique of size $\Delta + 1$. Denote the vertex set of the independent set, respectively the clique, by $B''$ and let $A'' \subseteq A'$ denote the set of vertices that are matched to vertices in $B''$. Then, $A'' \cup B''$ induces a $(\Delta + 2)$-net if $B''$ induces an independent set, and a $(\Delta + 1)$-prism if $B''$ induces a clique, contradicting the fact that $G$ is $\mathcal{H}_\Delta$-free. We conclude that $\omega(G[A]) \leq R(\Delta+2, \Delta+1) - 1$.
\end{proof}

Bui-Xuan, Telle, and Vatshelle \cite{BTV2013} showed that on graphs with bounded mim-width unweighted dominating set can be solved in polynomial time, provided that a branch decomposition of bounded mim-width is given as part of the input, and noted that the same framework can be used for the weighted case with only minor modifications and without affecting the running time. Since a linear ordering can be transformed in linear time into a caterpillar branch decomposition of the same mim-width, their result also yields a polynomial-time algorithm for the weighted dominating set problem given a linear ordering of bounded mim-width.

Thus, by \zcref{lem:dominating-forbidden-subgraphs, lem:dominating-lmimw}, instances of the weighted dominating set problem for which the coefficient matrix of~\eqref{eq:dominating-set-ip} is totally $\Delta$-modular can be solved in polynomial time for each fixed $\Delta$.

\section{Induced matching}
Given a graph $G=(V,E)$ and a weight function $w: E \to \Q$, the maximum induced matching problem asks for a subset $M \subseteq E$ of maximum weight such that $M$ is a matching and the graph induced by all endpoints of edges in $M$ has no edges other than $M$. There is a natural integer programming formulation of the maximum induced matching problem based on the \emph{line graph} $L(G)$ of $G$. Recall that $L(G)$ is the graph with vertex set $E$ in which $e, f \in E$ are adjacent if and only if they share a common endpoint in $G$. Then, the maximum induced matching problem has the following integer programming formulation:
\begin{equation}
    \label{eq:induced-matching-ip}
    \begin{aligned}
        \max\ & \sum_{e \in E} w(e) x_e \\
        \text{s.\,t.}\ & x_e + \sum_{f \in N_{L(G)}(e)} x_f \leq 1 && \qquad\forall e \in E\\
        & x_e \in \{0, 1\} && \qquad\forall e \in E
    \end{aligned}
\end{equation}

\begin{lemma}
    \label{lem:induced-matching}
    If the coefficient matrix of \eqref{eq:induced-matching-ip} is totally $\Delta$-modular, then the pathwidth of $G$ is at most $7 \lceil \log_2 \Delta \rceil + 5$.
\end{lemma}
\begin{proof}
    Let $H$ denote the \emph{1-subdivided claw}, i.e., the graph obtained from the claw by subdividing each edge exactly once and let $F$ denote the disjoint union of $\lceil \log_2 \Delta \rceil + 1$ copies of $H$. Note that $L(H)$ is a net and that if $G$ contains $F$ as a subgraph, then $L(G)$ contains the disjoint union of $\lceil \log_2 \Delta \rceil + 1$ nets as an induced subgraph. Moreover, observe that the coefficient matrix of \eqref{eq:induced-matching-ip} is the same as the coefficient matrix of \eqref{eq:dominating-set-ip} for the graph $L(G)$. By \zcref{lem:dominating-forbidden-subgraphs}, if the coefficient matrix of \eqref{eq:induced-matching-ip} is totally $\Delta$-modular, then $L(G)$ does not contain the disjoint union of $\lceil \log_2 \Delta \rceil + 1$ nets as an induced subgraph and hence, by the argument above, $G$ does not contain $F$ as a subgraph.

    Finally, since $F$ has maximum degree 3, if $G$ contains $F$ as a minor, then it also contains a subdivision of $F$ as a subgraph (see~\cite[Proposition 1.7.3]{diestel2025graph}). Since every subdivision of $H$ contains $H$ as a subgraph, we conclude that $G$ does not contain the forest $F$ as a minor. By a result of Bienstock, Robertson, Seymour, and Thomas \cite{bienstock1991quickly}, this implies that the pathwidth of $G$ is at most $|V(F)| - 2 = 7 \lceil \log_2 \Delta \rceil + 5$.
\end{proof}

By the algorithm of Dallard, Milani\v{c}, and \v{S}torgel~\cite[Corollary 7.4]{DMS2024} for weighted maximum independent packing,
the weighted maximum induced matching problem can be solved
in polynomial time on graphs of bounded pathwidth. We conclude that, for any fixed $\Delta$, totally $\Delta$-modular instances of the induced matching problem can be solved in polynomial time.

\section{Cluster vertex deletion}
\label{sec:clustervertexdel}
As already mentioned before, integer programs with at most two non-zero entries per row and bounded subdeterminants can be solved in polynomial time \cite{fiorini2025integer}. A natural follow-up question is whether this result can be extended to integer programs with a bounded number of non-zero entries per row. In this section, we will consider a problem which naturally falls into the case of at most three non-zero entries per row, namely the cluster vertex deletion problem.

Given a graph $G = (V, E)$ and a weight function $w: V \to \Q$, the cluster vertex deletion problem asks to find a subset $S \subseteq V$ of minimum weight such that $G - S$ is a cluster graph (a disjoint union of cliques). Here and in the following, $G - S$ denotes the induced subgraph of $G$ on the vertex set $V(G) \setminus S$. It is not difficult to see that a graph is a cluster graph if and only if it does not contain an induced $P_3$, where $P_m$ denotes the path on $m$ vertices. This yields the following integer programming formulation.
\begin{equation}
    \label{eq:cluster-vertex-del}
    \begin{aligned}
        \min\ & \sum_{v \in V} w(v) x_v \\
        \text{s.\,t.}\ & x_a + x_b + x_c \geq 1 && \qquad\forall \{a, b, c\} \in \mathcal{P} \\
        & x_v \in \{0, 1\} && \qquad\forall v \in V
    \end{aligned}
\end{equation}
where $\mathcal{P}$ is the set of all $\{a,b,c\} \subseteq V$ such that $G[\{a,b,c\}]$ is isomorphic to $P_3$, and $x_v$ indicates whether the vertex $v$ is included in the cluster vertex deletion set.

\begin{figure}[htbp]
    \centering
    \includegraphics{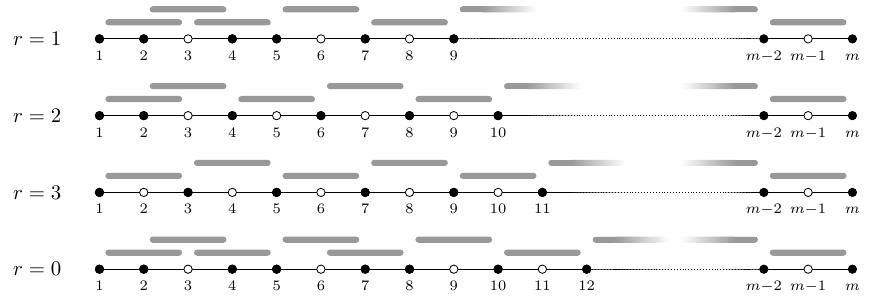}
    \caption{Illustration of the proof of \zcref{lem:cluster-vertex-del-chordal}. The shaded regions represent the selected induced $P_3$'s and the black vertices are the selected vertices.}
    \label{fig:cluster-vertex-del}
\end{figure}

An induced cycle of length at least four is called a \emph{hole}. A graph is \emph{chordal} if it does not contain any holes. The following lemma shows a connection between the cluster vertex deletion problem and chordal graphs.
\begin{lemma}
    \label{lem:cluster-vertex-del-chordal}
    There exists a computable function $f$ such that, if the coefficient matrix of \eqref{eq:cluster-vertex-del} is totally $\Delta$-modular, then there exists a set $X \subseteq V$ with $|X| \le f(\Delta)$ such that $G-X$ is chordal.
\end{lemma}
\begin{proof}
    We claim that any hole corresponds to a submatrix with determinant of absolute value at least 2 and that for multiple holes, they combine to form a block diagonal matrix. This implies that $G$ has at most $\lfloor\log_2 \Delta\rfloor$ vertex-disjoint holes and, by the Erdős-Pósa theorem for holes \cite{kim2020erdos,HK24}, we conclude that there is a set $X \subseteq V$ of size at most $f(\Delta)$ such that $G-X$ is chordal for some computable function $f$.

    It remains to show that any hole corresponds to a submatrix with determinant of absolute value at least 2. To this end, note that a sequence $r_0, \dots, r_{\ell-1}$ of induced $P_3$'s in the path $P_{m}$ with a sequence $c_0, c_1, \ldots, c_{\ell}$ of vertices such that $r_i$ contains exactly two of the $c_j$'s, namely $c_i$ and $c_{i+1}$, for all $i \in \{0, \ldots, \ell-1\}$ and where $c_0$ is the first vertex of the path and $c_{\ell}$ is the last vertex of the path gives rise to $\ell$ induced $P_3$'s in a hole of length $m-1$ and $\ell$ vertices in the hole by identifying $c_0$ and $c_{\ell}$. Moreover, the corresponding submatrix is the incidence matrix of a cycle of length $\ell$, which has determinant of absolute value 2 if $\ell$ is odd. Therefore, it suffices to show that for any $m \geq 5$, there exist sequences as described above for $P_{m}$ with $\ell$ odd. We can give these sequences explicitly depending on the residue $r$ of $m$ modulo 4 (see also the illustration in \zcref{fig:cluster-vertex-del}):
    \begin{alignat*}{3}
        r&=1\colon & &\{1,2,3\},\{2,3,4\},\{3,4,5\}, \text{ and }\{5,6,7\},\{7,8,9\},\dots, \{m-2,m-1,m\}; \\
        & & &1,2,4,5,\text{ and } 7, 9, \dots,m \\
        r&=2\colon & &\{1,2,3\},\{2,3,4\},\{4,5,6\}, \text{ and }\{6,7,8\},\{8,9,10\},\dots, \{m-2,m-1,m\}; \\
        & & & 1,2,4,6,\text{ and } 8, 10, \dots,m \\
        r&=3\colon & &\{1,2,3\},\{3,4,5\},\{5,6,7\}, \text{ and }\{7,8,9\},\{9,10,11\},\dots, \{m-2,m-1,m\};\\
        & & &1,3,5,7,\text{ and } 9, 11, \dots,m \\
        r&=0\colon & &\{1,2,3\},\{2,3,4\},\{3,4,5\},\{5,6,7\},\{6,7,8\},\text{ and }\{8,9,10\},\{10,11,12\},\dots,\\
        & & & \{m-2,m-1,m\}; \quad 1,2,4,5,7,8,\text{ and } 10, 12, \dots,m
    \end{alignat*}
    Here, each case starts with a construction for $5,6,7$ or $8$ vertices and then extends the construction to $m$ by appending pairs of induced $P_3$'s which intersect in their endpoints. Clearly, for pairwise vertex-disjoint holes the submatrices combine to form a block diagonal matrix.
\end{proof}

\begin{lemma}
    \label{lem:cluster-vertex-del}
    For any fixed $k$, if there exists a set $X$ of at most $k$ vertices such that $G-X$ is chordal, then the cluster vertex deletion problem can be reduced to a polynomial number of cluster vertex deletion problems on chordal graphs.
\end{lemma}
\begin{proof}
    By assumption there is a set $X$ of at most $k$ vertices such that $G-X$ is chordal. Since it is possible to recognize chordal graphs in linear time \cite{rose1976algorithmic}, we can find such a set $X$ in polynomial time.

    Our algorithm will work by guessing some part of the solution such that the remaining instance is chordal and such that there is a polynomial number of possible guesses. To this end, let $Y^*$ be an optimal solution and let $Y = Y^* \cap X$.

    Clearly, $Y$ hits all induced $P_3$'s that lie completely in $X$. Moreover, let $Z_Y$ be the set of vertices $z \in V \setminus X$ such that there is an induced $P_3$ with $z$ as an endpoint and the other two vertices in $X \setminus Y$. Then $Y^*$ must contain $Z_Y$. Further, let $N_x = N_G(x) \setminus X$ for each $x \in X \setminus Y$ where $N_G(x)$ is the neighborhood of $x$ in $G$. Observe that if $z_1,z_2 \in N_x$ are non-adjacent in $G$, then $z_1, x, z_2$ is an induced $P_3$ of $G$. Hence, $Y^*$ must contain $z_1$ or $z_2$. In other words, $Y^*$ contains a vertex cover of $\overline{G}[N_x]$ (and thus also an inclusion-wise minimal vertex cover), where $\overline{G}$ is the complement of $G$.

    We can therefore guess $Y$ and a minimal vertex cover $Z_x$ of $\overline{G}[N_x]$ for each $x \in X \setminus Y$. For each such guess we consider the remaining instance $G' = G - (Y \cup Z_Y \cup \bigcup_{x \in X \setminus Y} Z_x)$. For $Y = Y^* \cap X$, it holds that $G'$ is chordal: Clearly, $G' - X$ is chordal since it is an induced subgraph of $G-X$. Thus, if $G'$ contains a hole, it must contain a vertex $x \in X \cap V(G') = X \setminus Y$ with two non-adjacent neighbors $z_1, z_2 \in V(G')$. However, we cannot have $z_1, z_2 \in X$, since $Y$ hits all induced $P_3$'s that lie completely in $X$. Further, we cannot have $z_1, z_2 \in V(G') \setminus X$, since $Z_x$ is a vertex cover of $\overline{G}[N_x]$. Lastly, we cannot have that one of $z_1, z_2$ is in $X$ and the other in $V(G') \setminus X$, since then the other is in $Z_Y$. Consequently, $G'$ is chordal. We can therefore check for each guess whether the remaining instance $G'$ is chordal and, if so, solve the cluster vertex deletion problem on $G'$. For the correct guess of $Y$ and the corresponding $Z_x$, taking the union of the solution on $G'$ with $Y \cup Z_Y \cup \bigcup_{x \in X \setminus Y} Z_x$ yields an optimal solution. Thus, enumerating all guesses and returning the best solution among them yields an optimal solution for the original instance.

    There are only $2^k$ guesses for $Y$ since $|X| \leq k$. Moreover, note that a minimal vertex cover $Z_x$ of $\overline{G}[N_x]$ corresponds to a maximal clique $K$ of $G[N_x]$ via $Z_x = N_x \setminus K$. Further, since $G[N_x]$ is a chordal graph as an induced subgraph of $G - X$, $G[N_x]$ contains at most $|N_x|$ maximal cliques, and all maximal cliques of $G[N_x]$ can be enumerated efficiently \cite{gavril1972algorithms}. Thus, all minimal vertex covers of $\overline{G}[N_x]$ can be enumerated efficiently. Since we only have to guess $|X \setminus Y| \leq |X|$ many $Z_x$, we can enumerate all guesses jointly in polynomial time.
\end{proof}

By \zcref{lem:cluster-vertex-del-chordal, lem:cluster-vertex-del}, and a recent result of Cao and Li \cite{CaoLi2026} who show that the cluster vertex deletion problem can be solved in polynomial time on chordal graphs, we can solve the cluster vertex deletion problem in polynomial time on instances where the coefficient matrix of the formulation \eqref{eq:cluster-vertex-del} is totally $\Delta$-modular for any constant $\Delta$.

\section{Traveling salesman problem}
The traveling salesman problem asks to find a cycle of minimum cost that visits each vertex exactly once in a given graph $G = (V,E)$ with edge costs $c: E \to \Q$. Several prominent integer programming formulations can be found in the literature. First, we want to mention the formulation by Miller, Tucker, and Zemlin~\cite{miller1960integer}, which contains $|V|$ as a coefficient in the coefficient matrix. Similarly, the flow-based formulation by Gavish and Graves~\cite{gavish1978travelling} contains $|V|-1$ as a coefficient in the coefficient matrix. Therefore, these formulations are totally $\Delta$-modular only for graphs with at most $\Delta$, respectively $\Delta + 1$ vertices.

More interesting are the formulations proposed by Dantzig, Fulkerson, and Johnson~\cite{dantzig1954solution}. We denote by $\delta(S)$ the set of edges with exactly one endpoint in $S$ and define $\delta(v) = \delta(\{v\})$. Both formulations share the constraints
\begin{equation}
    \label{eq:tsp-dfj-both}
    \begin{aligned}
        \min\ & \sum_{e \in E} c(e) x_e \\
        \text{s.\,t.}\ & \sum_{e \in \delta(v)} x_e = 2 && \qquad\forall v \in V \\
        & x_e \in \{0, 1\} && \qquad\forall e \in E
    \end{aligned}
\end{equation}
where $x_e$ indicates whether the edge $e$ is included in the cycle.
The first formulation has \eqref{eq:tsp-dfj-both} and the additional constraints
\begin{alignat}{2}
    \label{eq:tsp-dfj-cut}
    & \sum_{e \in \delta(S)} x_e \geq 2 && \alignatquad\forall \emptyset \neq S \subsetneq V,
\end{alignat}
the second has \eqref{eq:tsp-dfj-both} and the additional constraints
\begin{alignat}{2}
    \label{eq:tsp-dfj-conn}
    & \sum_{e \in E(S)} x_e \leq |S| - 1 && \alignatquad\forall \emptyset \neq S \subsetneq V,
\end{alignat}
where $E(S)$ denotes the set of edges with both endpoints in $S$.

Note that \eqref{eq:tsp-dfj-cut} and \eqref{eq:tsp-dfj-conn} are equivalent in the sense that one arises from the other by adding, respectively subtracting, some constraints of \eqref{eq:tsp-dfj-both}, and dividing by 2. However, this operation does not preserve subdeterminants, and therefore the two formulations are not directly equivalent in the context of total $\Delta$-modularity.

\begin{lemma}
    \label{lem:tsp-dfj}
    If the coefficient matrix of \eqref{eq:tsp-dfj-both} and \eqref{eq:tsp-dfj-cut}, respectively of \eqref{eq:tsp-dfj-both} and \eqref{eq:tsp-dfj-conn}, is totally $\Delta$-modular, then any cycle in $G$ has length at most $\max\{4, 4\log_2 \Delta + 3\}$.
\end{lemma}
\begin{proof}
    Assume $G$ contains a cycle $C$ of length $l \geq 5$. Denote the vertices of the cycle by $V(C) = \{v_0, \ldots, v_{\ell-1}\}$ and the edges by $e_i = \{v_i, v_{i+1}\}$ for $i \in \{0, \ldots, \ell-1\}$, where the indices are taken modulo $\ell$.
    
    Consider the submatrix that selects the columns corresponding to the edges $e_a, e_{a+1}, e_{a+3}$ for $a \in \{0,4,\ldots,4 (\lfloor \ell/4 \rfloor - 1)\}$ and three rows for each such $a$. We claim that for a suitable choice of the rows, the resulting submatrix is a block-diagonal matrix with one block of determinant $2$ for each $a$. First, select the row from \eqref{eq:tsp-dfj-both} corresponding to the degree constraint of $v_{a+1}$, then for \eqref{eq:tsp-dfj-cut} additionally select the rows corresponding to $S = \{v_{a+2}, v_{a+3}\}$ and to $S = \{v_{a+1}, v_{a+2}, v_{a+3}\}$, respectively for \eqref{eq:tsp-dfj-conn} additionally select the rows corresponding to $S = \{v_{a+1}, \dots, v_{a+4}\}$ and to $S = \{v_a, v_{a+1}, v_{a+3}, v_{a+4}\}$. Since $\ell \geq 5$, the rows selected for $a$ have support in $e_a, e_{a+1}, e_{a+3}$ when restricted to the selected columns, so the resulting submatrix is block-diagonal and each block is equal to $
        \begin{psmallmatrix}
            1 & 1 & 0 \\
            0 & 1 & 1 \\
            1 & 0 & 1
        \end{psmallmatrix}
    $ which has determinant $2$. We conclude that $2^{\lfloor \ell/4 \rfloor} \leq \Delta$ which implies that $\ell \leq 4 \log_2 \Delta + 3$.
\end{proof}

By \zcref{lem:tsp-dfj}, any feasible instance of the traveling salesman problem with a totally $\Delta$-modular coefficient matrix with regard to either of the formulations by Dantzig, Fulkerson, and Johnson has at most $\max\{4, 4\log_2 \Delta + 3\}$ vertices. Therefore, for any fixed $\Delta$, the traveling salesman problem can be solved in polynomial time by brute force on such instances.

Finally, Claus \cite{claus1984new} proposed a formulation which is based on the idea of considering the directed graph $D = (V, A)$ obtained from $G$ by replacing each edge $\{u, v\} \in E$ by the two arcs $(u, v), (v, u) \in A$, defining the cost of each arc $(u, v) \in A$ as $c(u, v) = c(v, u) = c(\{u, v\})$, and finding a directed cycle of minimum cost that visits each vertex exactly once in $D$. The formulation selects an arbitrary vertex $s \in V$ as a source and is given by
\begin{equation}
    \label{eq:tsp-flow-ip}
    \begin{aligned}
        \min\ & \sum_{a \in A} c(a) x_a \\
        \text{s.\,t.}\ & \sum_{a \in \delta^+(v)} x_a = 1 && \qquad\forall v \in V \\
        & \sum_{a \in \delta^-(v)} x_a = 1 && \qquad\forall v \in V \\
        & y_a^{t} - x_a \leq 0 && \qquad\forall a \in A, t \in V \setminus \{s\} \\
        & \sum_{a \in \delta^+(v)} y_a^{t} - \sum_{a \in \delta^-(v)} y_a^{t} = b_v^{t} && \qquad\forall v \in V, t \in V \setminus \{s\} \\
        & x_a \in \{0, 1\} && \qquad\forall a \in A \\
        & y_a^{t} \in \{0, 1\} && \qquad\forall a \in A, t \in V \setminus \{s\}
    \end{aligned}
\end{equation}
where $b_s^{t} = 1$, $b_t^{t} = -1$, and $b_v^{t} = 0$ for all $v \in V \setminus \{s, t\}$, and $\delta^+(v)$ and $\delta^-(v)$ denote the sets of outgoing and incoming arcs of $v$, respectively. In the formulation, $x_a$ indicates whether the arc $a$ is included in the cycle and $f_t: A \to \{0,1\}, f_t(a) = y_a^{t}$ is a unit flow from $s$ to $t$ for each $t \in V \setminus \{s\}$.

\begin{lemma}
    \label{lem:tsp-flow}
    If the coefficient matrix of \eqref{eq:tsp-flow-ip} is totally $\Delta$-modular, then any cycle in $G$ has length at most $2 \log_2 \Delta + 1$.
\end{lemma}
\begin{proof}
    We refer to the constraint families (in the order they appear in \eqref{eq:tsp-flow-ip}) as the outdegree constraints, the indegree constraints, the coupling constraints, and the flow conservation constraints.

    Assume $G$ contains a cycle $C$ of length $\ell$. Label the vertices of the cycle by $v_0, \ldots, v_{\ell-1}$ in cyclic order. Since $A$ contains both orientations of every edge of $G$, the arcs $(v_i, v_{i+1})$ and $(v_{i+1}, v_i)$ are contained in $A$ for all $i$. Moreover, since $|V| \geq \ell \geq 3$, we can fix two distinct vertices $t_1, t_2 \in V \setminus \{s\}$.

    Consider the submatrix that selects the columns corresponding to the variables $x_{(v_{i-1}, v_i)}$, $x_{(v_i, v_{i+1})}$, $x_{(v_i, v_{i-1})}$, $y^{t_1}_{(v_{i-1}, v_i)}$, $y^{t_1}_{(v_i, v_{i+1})}$, $y^{t_2}_{(v_{i-1}, v_i)}$, $y^{t_2}_{(v_i, v_{i-1})}$ for $i \in \{0, 2, \ldots, 2 (\lfloor \ell/2 \rfloor - 1)\}$ and seven rows for each such $i$. We claim that for a suitable choice of the rows, the resulting submatrix is a block-diagonal matrix with one block of determinant $2$ for each $i$. For each $i$, select the row corresponding to the outdegree constraint of $v_i$, the rows corresponding to the coupling constraints for $((v_{i-1}, v_i), t_1)$, $((v_i, v_{i+1}), t_1)$, $((v_{i-1}, v_i), t_2)$ and $((v_i, v_{i-1}), t_2)$, and the rows corresponding to the flow conservation constraints for $(v_i, t_1)$ and $(v_i, t_2)$. Each of these rows has nonzero entries only in variables of arcs incident to $v_i$, and each column selected for $i$ corresponds to an arc of $C$ incident to $v_i$. Since no two of the vertices $v_i$ are adjacent in $C$, this means the resulting submatrix is block-diagonal and, with rows and columns ordered as listed above, each block is equal to
    \begin{equation*}
        \begin{pmatrix}
            0  & 1  & 1  & 0  & 0 & 0  & 0 \\
            -1 & 0  & 0  & 1  & 0 & 0  & 0 \\
            0  & -1 & 0  & 0  & 1 & 0  & 0 \\
            -1 & 0  & 0  & 0  & 0 & 1  & 0 \\
            0  & 0  & -1 & 0  & 0 & 0  & 1 \\
            0  & 0  & 0  & -1 & 1 & 0  & 0 \\
            0  & 0  & 0  & 0  & 0 & -1 & 1
        \end{pmatrix},
    \end{equation*}
    which has determinant $2$. We conclude that $2^{\lfloor \ell/2 \rfloor} \leq \Delta$ which implies that $\ell \leq 2 \log_2 \Delta + 1$.
\end{proof}

By \zcref{lem:tsp-flow}, any feasible instance of the traveling salesman problem with a totally $\Delta$-modular coefficient matrix with regard to \eqref{eq:tsp-flow-ip} has at most $2 \log_2 \Delta + 1$ vertices and can thus be solved in polynomial time by brute force.

\section{2-constraint satisfaction problem}
We consider the Boolean version of the 2-constraint satisfaction problem in the weighted optimization variant (max 2-CSP).
An instance consists of a finite set of variables $X$, a finite set $C$ of clauses, each of the form $((x, y), P)$ with distinct $x, y \in X$ and a predicate $P: \{0, 1\}^2 \to \{0, 1\}$, and a weight function $w: C \to \Q_{\geq 0}$. The problem then asks for an assignment $\sigma: X \to \{0, 1\}$ that maximizes the total weight of the satisfied clauses, where a clause $((x, y), P)$ is satisfied by $\sigma$ if $P(\sigma(x), \sigma(y)) = 1$.

For a predicate $P$, we define $Q_P = \conv\{(x,y,z) \in \{0, 1\}^3 : z \leq P(x, y)\}$. Up to replacing $x$ by $(1-x)$ or $y$ by $(1-y)$, every predicate is of one of the following four types: AND ($P(x,y)=x\land y$), OR ($P(x,y)=x\lor y$), XOR ($P(x,y)=x\oplus y$) or affine ($P(x,y)$ is an affine function over $\R$ of $x$ and $y$). Note that in the affine case, $P$ is either constant or equal to $x$ or $y$, up to negation.
For each of these, we can give a description of $Q_P$ by linear inequalities, which are given by the bounds $0 \leq x, y, z \leq 1$ and
\begin{equation}
    \label{eq:2csp-rel-ineqs}
    \begin{aligned}
        \text{AND: }& z - x \leq 0, \quad z - y \leq 0, &\qquad \text{OR: }& z - x - y \leq 0, \\
        \text{XOR: }& z - x - y \leq 0, \quad z + x + y \leq 2, &\qquad\text{affine: }& z - P(x,y) \leq 0.
    \end{aligned}
\end{equation}

With this notation, max 2-CSP can be formulated as the following integer program, where the containment in $Q_P$ is expressed by the inequalities above, with $(1-x)$ and $(1-y)$ substituted for $x$ and $y$ appropriately.
\begin{equation}
    \label{eq:2csp-ip}
    \begin{aligned}
        \max\ & \sum_{c \in C} w(c) z_c \\
        \text{s.\,t.}\ & (x, y, z_c) \in Q_P && \qquad\forall c = ((x, y), P) \in C \\
        & x \in \{0, 1\} && \qquad\forall x \in X \\
        & z_c \in \{0, 1\} && \qquad\forall c \in C
    \end{aligned}
\end{equation}

Max 2-CSP includes max 2-SAT, max cut, and directed max cut as special cases. Further, the formulation \eqref{eq:2csp-ip} generalizes these problems on the level of integer programming, i.e., specializing it to the respective problems yields a natural integer programming formulation for each of them.

For a first structural result, we examine the set $X_\oplus \subseteq X$ of variables that appear in XOR clauses. To this end, let $G$ be the (multi)graph with vertex set $X_\oplus$ that has an edge $\{x, y\}$ for each XOR clause $((x, y), P) \in C$, where we identify each edge with the corresponding clause. Note that, as in the classification above, XOR clauses include clauses in which $x$ or $y$ appear negated.

\begin{lemma}
    \label{lem:2csp-xor}
    If the coefficient matrix of \eqref{eq:2csp-ip} is totally $\Delta$-modular, then $|X_\oplus| \leq 2 \log_2\Delta$.
\end{lemma}
\begin{proof}
    First, consider the case that $G$ is connected and let $T$ be a spanning tree of $G$. Let $v_1, \dots, v_n$ be an ordering of the vertices of $T$ and let $e_1, \dots, e_{n-1}$ be an ordering of the edges of $T$ such that $T_i = (\{v_1, \dots, v_i\}, \{e_1, \dots, e_{i-1}\})$ is a connected subtree of $T$ for all $2 \le i \le n$. We now iteratively construct a submatrix $A_i$ of the coefficient matrix with $|\det(A_i)| = 2^{i-1}$ for $i = 2, \dots, n$.

    For $i = 2$, let $A_2$ be the submatrix whose rows correspond to the two constraints in \eqref{eq:2csp-rel-ineqs} of the clause $e_1 = \{v_1, v_2\}$ and whose columns correspond to the variables $z_{e_1}$ and $v_2$. Then, up to flipping the signs in the second column if $v_2$ appears negated in $e_1$, we have $A_2 = \begin{psmallmatrix} 1 & -1 \\ 1 & 1 \end{psmallmatrix}$, and, in any case, $|\det(A_2)| = 2$.

    For $i > 2$, let $A_i$ arise from $A_{i-1}$ by additionally selecting the rows corresponding to the two constraints of the clause $e_{i-1}$ and the columns corresponding to the variables $z_{e_{i-1}}$ and $v_i$. Note that the new columns are zero in all rows of $A_{i-1}$ and that $v_i \in e_{i-1}$. Therefore, $A_i$ has the block lower-triangular form
    \[
        A_i = \begin{pmatrix} A_{i-1} & \zero \\ B & C \end{pmatrix}, \quad C = \begin{pmatrix} 1 & -1 \\ 1 & 1 \end{pmatrix},
    \]
    where, as before, the signs in the second column of $C$ are flipped if $v_i$ appears negated in $e_{i-1}$, and $|\det(A_i)| = |\det(A_{i-1})|\cdot|\det(C)| = 2^{i-2}\cdot 2 = 2^{i-1}$.

    In general, each connected component can be handled separately in the way described above. Since the resulting submatrices use disjoint rows and columns, combining them yields a block diagonal matrix whose determinant is the product of the determinants of the blocks. If these components have $n_1, \dots, n_k$ vertices, this yields $2^{(n_1 - 1) + \dots + (n_k - 1)} \leq \Delta$. Since each connected component has at least one edge, we have $n_j - 1 \geq n_j/2$ for each $j$, and the claim follows.
\end{proof}

\begin{lemma}
    \label{lem:2csp}
    For every instance of max 2-CSP and every assignment $\tau\colon X_\oplus \to \{0, 1\}$, the integer program obtained from \eqref{eq:2csp-ip} by fixing the variables in $X_\oplus$ according to $\tau$ is equivalent to an integer program with at most two nonzero entries per row. Moreover, this integer program can be computed in linear time and the largest absolute value of a subdeterminant of its coefficient matrix is at most that of the coefficient matrix of \eqref{eq:2csp-ip}.
\end{lemma}
\begin{proof}
    Fixing the variables in $X_\oplus$ corresponds to deleting their columns from the coefficient matrix of \eqref{eq:2csp-ip} and adjusting the right-hand side, so every subdeterminant of the resulting coefficient matrix is a subdeterminant of the coefficient matrix of \eqref{eq:2csp-ip}. Since both variables of every XOR clause are fixed, by \eqref{eq:2csp-rel-ineqs}, the only rows of the resulting coefficient matrix that have more than two nonzero entries are those corresponding to OR clauses $((x, y), P)$ with $x, y \notin X_\oplus$. For each such clause $c = ((x, y), P)$, we introduce two new variables $z_{c,x}$ and $z_{c,y}$ that indicate which of the two variables is responsible for satisfying the clause. Then, we substitute $z_{c,x} + z_{c,y}$ for $z_c$ in the objective and replace the constraints in \eqref{eq:2csp-rel-ineqs} by the following constraints, where $x$ is replaced by $(1-x)$ or $y$ by $(1-y)$ consistently:
    \begin{equation}
        \label{eq:2csp-gadget}
        z_{c,x} - x \leq 0, \quad z_{c,y} - y \leq 0, \quad z_{c,x} + z_{c,y} \leq 1, \quad z_{c,x}, z_{c,y} \in \{0, 1\}.
    \end{equation}
    For fixed $x, y \in \{0, 1\}$, the feasible values of $z_{c,x} + z_{c,y}$ under these constraints are precisely the values $z_c \in \{0, 1\}$ with $z_c \leq P(x, y)$. Thus, the resulting integer program is equivalent to the previous one via $z_c = z_{c,x} + z_{c,y}$. Clearly, its coefficient matrix has at most two nonzero entries per row and the transformation can be computed in linear time.

    Moreover, we claim that the largest absolute value of a subdeterminant of the coefficient matrix does not increase by this transformation. To this end, consider the coefficient matrices without the rows corresponding to the bounds on the variables, which do not affect the largest absolute value of a subdeterminant. Then the effect of the transformation for a single OR clause is to replace
    \[
        \bordermatrix{
            & z_c & x & y & \cr
            & 1 & -1 & -1 & \zero \cr
            & \zero & \ast & \ast & \ast
        }
        \quad \text{by} \quad
        \bordermatrix{
            & z_{c,x} & z_{c,y} & x & y & \cr
            & 1 & 0 & -1 & 0 & \zero \cr
            & 0 & 1 & 0 & -1 & \zero \cr
            & 1 & 1 & 0 & 0 & \zero \cr
            & \zero & \zero & \ast & \ast & \ast
        },
    \]
    where the columns are labeled by the corresponding variables, the unlabeled columns correspond to the remaining variables, and the entries marked by $\ast$ denote submatrices that remain unchanged. The rows appear in the order of appearance in \eqref{eq:2csp-gadget} with the remaining rows below. If $x$ is replaced by $1-x$ or $y$ by $1-y$ in the clause, the signs of the corresponding entries are flipped.

    It remains to show that for every square submatrix $M'$ of the transformed matrix with $\det(M') \neq 0$, there is a square submatrix $M$ of the original matrix with $|\det(M)| = |\det(M')|$. Let $r_1, r_2, r_3$ denote the first three rows of the transformed matrix and $r$ the first row of the original matrix, and call $r_1, r_2, r_3$ and the columns of $z_{c,x}$ and $z_{c,y}$ \emph{new}. As long as a new row or column has exactly one nonzero entry in $M'$, we expand the determinant along it and replace $M'$ by the resulting smaller square submatrix of the transformed matrix. Since this entry is $\pm 1$, the absolute value of the determinant does not change. Afterwards, every new row and column in $M'$ has at least two nonzero entries in $M'$. If $M'$ contains no new column, it thus contains no new row either, as each new row has at most one nonzero entry outside the new columns, so $M'$ is a submatrix of the original matrix. Otherwise, $M'$ contains a new column and hence both rows in which this column is nonzero, including $r_3$. As $r_3$ is nonzero only in the new columns, $M'$ contains both new columns and hence all new rows. Replacing $r_3$ by $r_3 - r_1 - r_2$ does not change $\det(M')$ and, regardless of the signs of the entries for $x$ and $y$, yields a row that is zero in the new columns and coincides with $-r$ in all other columns. Expanding along the two new columns, which now have exactly one nonzero entry each, shows that $|\det(M')| = |\det(M)|$, where $M$ is the submatrix of the original matrix whose rows are $r$ and the rows of $M'$ other than the new ones, and whose columns are those of $M'$ other than the new ones.

    Since the OR clauses can be transformed one at a time and the above argument does not depend on the unchanged part of the matrix, this proves the claim.
\end{proof}

If the coefficient matrix of \eqref{eq:2csp-ip} is totally $\Delta$-modular, then $|X_\oplus| \leq 2 \log_2\Delta$ by \zcref{lem:2csp-xor}, so there are at most $\Delta^2$ assignments $\tau\colon X_\oplus \to \{0, 1\}$. Hence, by \zcref{lem:2csp} and the result of \cite{fiorini2025integer}, for any fixed $\Delta$, max 2-CSP can be solved in polynomial time on instances where the coefficient matrix of \eqref{eq:2csp-ip} is totally $\Delta$-modular by solving the integer program from \zcref{lem:2csp} for each such assignment and taking the best solution.

As mentioned before, this applies in particular to max 2-SAT, max cut, and directed max cut as special cases of max 2-CSP.
For directed max cut, all clauses are AND clauses, so the coefficient matrix of \eqref{eq:2csp-ip} already has at most two nonzero entries per row and the result of \cite{fiorini2025integer} applies directly. For max cut, all clauses are XOR clauses and there is an XOR clause for each edge. Hence, by \zcref{lem:2csp-xor}, instances where the coefficient matrix of \eqref{eq:2csp-ip} is totally $\Delta$-modular have at most $2 \log_2\Delta$ non-isolated vertices and can be solved by enumerating the at most $\Delta^2$ assignments of these vertices. Finally, for max 2-SAT, all clauses are OR or affine clauses and the transformation from \zcref{lem:2csp} is needed to apply the result of \cite{fiorini2025integer}.

\section*{Acknowledgements}
JPG was supported in part by the Slovenian Research and Innovation Agency (research project N1-0370).
TH, OK, and SO were supported by the Institute for Basic Science (IBS-R029-C1).
OK was also supported by the National Research Foundation of Korea (NRF) grant funded by
the Ministry of Science and ICT (No. RS-2023-00211670).
RS was supported by the Young Scientist Fellowship of the Institute for Basic Science (IBS-R029-Y8).

\printbibliography

\end{document}